\documentstyle[eqsecnum,aps,epsf]{revtex}
\begin{document}
\title{SPIN EFFECTS IN GRAVITATIONAL RADIATION BACKREACTION\hskip5cm
III. COMPACT BINARIES WITH TWO SPINNING COMPONENTS}
\author{L\'aszl\'o \'A. Gergely, Zolt\'an I. Perj\'es and M\'aty\'as Vas\'uth}
\address{KFKI Research Institute for Particle and Nuclear\\
Physics, Budapest 114, P.O.Box 49, H-1525 Hungary}
\maketitle
 
\begin{abstract}
 
The secular evolution of a spinning, massive binary system in
eccentric orbit is analyzed, expanding and
generalizing our previous treatments of the Lense-Thirring motion and
the one-spin limit. The spin-orbit and spin-spin effects up to
the 3/2 post-Newtonian order are considered,
both in the equations of motion and in the radiative losses.
The description of the orbit in terms
of the true anomaly parametrization provides a simple averaging
technique, based on the residue theorem, over eccentric orbits. The
evolution equations of the angle variables characterizing the relative
orientation of the spin and orbital angular momenta reveal a speed-up
effect due to the eccentricity.
The dissipative evolution of the relevant dynamical and angular
variables is presented in the form of a closed system of
differential equations.
 
\end{abstract}
 
\pacs{PACS Numbers: }
 
\section{Introduction}
 
The motion of a binary system of two spinning bodies under the influence of
gravitational radiation has been treated in several recent works.
 The aim of these investigations is to obtain a self-consistent
description of the evolution, such that templates for the gravitational
radiation pattern can be provided for the wave observatories under
construction. Spin effects modulate both the amplitude and the frequency
of the waves. As of now, the basic properties of the radiative
evolution have been understood in a perturbative framework, both in a
post-Newtonian expansion\cite{KWW}-\cite{GPV2} and by black-hole
perturbation techniques\cite{MST}-\cite{OTO}. The post-Newtonian
expansion proceeds in powers of the parameter $\epsilon\approx
v^2/c^2\approx Gm/c^2r$.
In both approaches, the radiative losses in characteristic quantities,
with the inclusion of spin-orbit and spin-spin
effects, have been computed. The averages over
circular orbits have been obtained in \cite{KWW} and \cite{Kidder}.
 
The importance of eccentric orbits in various physical
scenarios has been emphasized by several authors\cite{GI}-\cite{HiBe}.
Quinlan and Shapiro\cite{QuSha} argue that clusters in galactic nuclei
in the final stage of collapse will contain a significant number of
eccentric binaries. Hills and Bender\cite{HiBe} suggest that many massive
$(M\approx 10^6-10^7 M_{\odot})$
 compact objects in the galactic centers
are gravitationally deflected by others and there is insufficient time
left for circularization before plunging. The behavior of eccentric
binaries under radiation reaction forces
has been studied in \cite{KWW}, \cite{Kidder} and \cite{WW}, and
(neglecting spin effects) in \cite{GI}.  Averaged radiative losses
for eccentric orbits have been obtained for a test particle by Ryan
\cite{Ry}. For finite masses, partial
descriptions (by computing the radiation losses of $E$ and $L$) have
been given by Rieth and Sch\"afer \cite{RS}.
The averaging procedure for eccentric orbits
has proven cumbersome in these works.
 
  In the present series of papers, we investigate the influence of the
intrinsic spin on the evolution of a radiating eccentric binary system
in a post-Newtonian approach.
The presence of spins complicates the description of the orbit
considerably; indeed the computations neglecting spins
have reached precisions up to the 7/2 PN order
\cite{Blanchet0A}-\cite{Blanchet0B}.
In papers \cite{GPV1} and
\cite{GPV2} (to be referred to as {\bf I} and {\bf II}, respectively) we
have chiseled two convenient
tools for our investigation. The first is the parametrization of
the orbit by the generalized {\em eccentric anomaly} ($\xi$) and the
{\em true anomaly} ($\chi$) parameters.
These parametrizations have already been employed in various special
cases. The test-particle limit has been considered in {\bf I}
and the one-spin limit in {\bf II}.
The second ingredient of our technology is the use of the residue theorem
for averaging the gravitational radiation losses by means of
these parametrizations. We now further develop and employ both
of these tools for the treatment of two spinning masses.
 
 For the purpose of predicting the evolution of a binary with two spins,
it is crucial
to determine the variation of a complete set of geometrical parameters characterizing
not just the orbit but also the orientation of the angular momenta.
The amplitude and polarization
angle of the gravitational signal is modulated
(as has been stressed in \cite{ACST}) by the changing
orientation of the binary in the observer's frame of reference.
 
  In this paper we characterize the
motion by computing the evolution of the system parameters to the 3/2 PN
order, including spin-orbit and (as an order-of-magnitude
analysis will reveal) spin-spin contributions.
 
The plan of the paper is as follows. In Sec. II., the motion
of the binary system is described following \cite{BOC}, \cite{TH}
and \cite{KWW}. We use the covariant spin supplementary condition
({\em SSC}) \cite{Kidder}. Then we introduce the
conserved quantities of the nonradiative motion: the energy $E$ and
the magnitude of orbital angular momentum $L$, further
the angles subtended by the orbital angular momenta and spins.
The evolution of these angles has been discussed for circular orbits in
\cite{ACST}, \cite{Apost} and \cite{deFelice}.
Here we give both the instantaneous and averaged evolution equations of
these angles, to 3/2 PN order, including the relevant spin-spin terms,
for eccentric orbits.
 
In Sec. III we compute the instantaneous losses of the energy $E$,
magnitude $L$ of the orbital angular momentum and the angle
variables from
Kidder's universal expressions (in \cite{Kidder}), and by use of the
Burke-Thorne \cite{BT} potential.
 
 We derive the secular evolution of the system variables due
to gravitational radiation reaction (Sec. IV). The contributions from
the Burke-Thorne potential have no effect on the averages over a period
of revolution. Our results provide a dual description [Eqs.
(\ref{avelossE})-(\ref{avelosska})
in terms of the energy $E$, orbital angular momentum $L$ and the spin
angles  {\it vs.} Eqs.
(\ref{avelosse})-(\ref{avelosska1}) in terms of the semimajor axis
$a$, eccentricity $e$ and the spin angles]
of the radiating binary system. The results
in {\bf I} and {\bf II} are special limiting cases.
 
\section{The Orbit of the Binary System}
 
The bound state of a two-body system with masses $m_1$ and $m_2$ and spins
${\bf S_1}$ and ${\bf S_2}$, respectively, is described by the Lagrangian
\begin{equation}
{\cal L}=\frac{\mu{\bf v}^2}{2}+{\frac{G m \mu}{r}}+
{\frac{2G\mu}{c^2r^3}}{\bf v}\cdot [{\bf r}\times ({\bf S}+
{\mbox{\boldmath $\sigma$}})] +{\frac{\mu}{2c^2m}}{\bf v}
\cdot ({\bf a}\times {\mbox{\boldmath $\sigma$}}) \ .
\label{delag}
\end{equation}
Here $r=\vert{\bf r}\vert$ is the relative distance, ${\bf v}$ the
relative velocity,
$\mu$ the reduced mass and $m$ the total mass of the system,
\begin{equation}
\mu={\frac{m_1 m_2}{m_1+m_2}}\ ,\qquad m=m_1+m_2\ ,
\qquad \eta={\frac{m_2}{m_1}} \ .
\end{equation}
The total and weighted spins are defined
\begin{equation}
{\bf S}={\bf S_1+S_2}\ ,\qquad
\mbox{\boldmath $\sigma$}=\eta{\bf S_1}+\eta^{-1}{\bf S_2}
\ .
\end{equation}
We have kept only the leading-order spin-orbit coupling, and adopted the
spin supplementary condition of\cite{Kidder}. The relative acceleration
entering the Lagrangian is
\begin{equation}
{\bf a}=-{\frac{Gm}{r^3}}{\bf r}+{\frac{G}{c^2r^3}} \left\{{\frac{6}{r^2}}
{\bf r} \left[({\bf r}\times{\bf v})\cdot({\bf S}+\mbox{\boldmath $\sigma$})\right] -
{\bf v\times(}4{\bf S}+3{\mbox{\boldmath $\sigma$})} +3{\frac{\dot r}{r}}{\bf r}\times (2
 {\bf S}+{\mbox{\boldmath $\sigma$}}) \right\}\ ,  \label{acc}
\end{equation}
and the momenta:
\begin{equation}
{\bf q}={\frac{\partial {\cal L}}{\partial{\bf a}}} ={\frac{\mu}{2c^2m}}
{\mbox {\boldmath $\sigma$}}\times {\bf v} \ ,  \label{q}
\end{equation}
\begin{equation}
{\bf p}={\frac{\partial {\cal L}}{\partial {\bf v}}}-{\bf \dot q} =\mu {\bf v
}+{\frac{G\mu}{c^2r^3}}{\bf r}\times(2{\bf
S}+{\mbox{\boldmath$\sigma$}}) \
. \label{p} \end{equation}
Here an overdot denotes derivative with respect to the time parameter $t$.
The constants of motion are the energy
$E={\bf p}\cdot {\bf v}+{\bf q}\cdot {\bf a}-{\cal L} $ and the total
angular momentum
${\bf J}={\bf S}+{\bf L}$.
The orbital angular momentum is
\begin{equation}
{\bf L}={\bf r}\times {\bf p}+{\bf v}\times {\bf q} ={\bf L_N}+{\bf L_{SO}}\
,  \label{Lvect}
\end{equation}
with the Newtonian orbital angular momentum and spin-orbit terms
\begin{equation}
{\bf L_N}=\mu {\bf r}\times{\bf v}\ ,
\end{equation}
\begin{equation}  \label{LSO}
{\bf L_{SO}}={\frac{\mu}{c^2m}} \left\{{\frac{Gm}{r^3}} \left[{\bf r}\times (
{\bf r}\times(2{\bf S}+{\mbox{\boldmath $\sigma$}}))\right]
-{\frac{1}{2}}\left[{\bf
v}\times ( {\bf v}\times\mbox{\boldmath $\sigma$})\right] \right\}\ .
\end{equation}
 
The spin precession equations have been given in \cite{Kidder}:
\begin{eqnarray}
{\bf \dot S_1}&=&{\frac{G}{c^2r^3}}\left(\frac{4+3\eta}{2}{\bf L_N - S_2} +
\frac{3}{r^2}({\bf r\cdot S_2}){\bf r}\right) \times{\bf S_1}\ ,  \nonumber
\\
{\bf \dot S_2}&=&{\frac{G}{c^2r^3}}\left(\frac{4+3\eta^{-1}}{2}{\bf L_N - S_1
} + \frac{3}{r^2}({\bf r\cdot S_1}){\bf r}\right) \times{\bf S_2}\ \ .
\label{Sprec}
\end{eqnarray}
The magnitudes $S_1$ and $S_2$ of both spins are separately constants.
 
All spin effects are formally of 1 PN order. A simple
estimate $S/L\approx\epsilon^{1/2}$ reveals however that the spin-orbit
and spin-spin terms are of respective orders
$\epsilon^{3/2}$ and $\epsilon^{2}$ in general.
Exceptionally, in the spin precession Eqs. (\ref{Sprec}) (predicting the
evolution of the {\em directions} ${\bf \hat S_i}$ of spins),
the spin-orbit and spin-spin contributions are of
order $\epsilon$ and $\epsilon^{3/2}$.
 
The precession rate of the orbital angular momentum, however,
will not include spin-spin terms, as they are of $\epsilon^{2}$
order. From the conservation of the total angular momentum ${\bf J}$
and from (\ref{Sprec} ) we obtain
\begin{equation}
{\bf \dot L}=-{\bf \dot S}={\frac{G}{2c^2r^3}} (4{\bf S}+3\mbox{\boldmath $\sigma$})\times
{\bf L}\ .  \label{Lprec}
\end{equation}
We substituted here, as we did in {\bf II}, ${\bf L}$ in place of the
Newtonian
angular momentum ${\bf L_N}$. It follows from (\ref{Lprec}) that the
magnitude $L$ of the orbital momentum and its projection on the vector $4
{\bf S}+3\mbox{\boldmath $\sigma$}$ are conserved. However, the total spin ${\bf S}$
shows a more complicated motion pattern.
 
The energy and the orbital momentum square are
\begin{eqnarray}
E &=&{\frac{\mu v^{2}}{2}}-{\frac{Gm\mu }{r}}+{\frac{G({\mbox{\boldmath
{\bf L}$\cdot \sigma$} })
}{c^{2}r^{3}}}\ ={\frac{\mu }{2}[\dot{r}^{2}+r^{2}(\dot{\theta}^{2}+\sin
^{2}\theta \ \dot{\varphi}^{2})]}-{\frac{Gm\mu }{r}}+{\frac{G(
{\mbox{\boldmath {\bf L}$\cdot \sigma$}  })}{c^{2}r^{3}}}\ ,  \label{E} \\
L^{2} &=&{\mu ^{2}}r^{4}(\dot{\theta}^{2}+\sin ^{2}\theta \ \dot{\varphi}
^{2})-4{\frac{G\mu ({\bf L\cdot S})}{c^{2}r}}+{\frac{2E({\mbox{\boldmath
{\bf L}$\cdot \sigma$} }) }{c^{2}m}}\ .  \label{L2}
\end{eqnarray}
Hence we express $v^{2}$ and $\dot{r}^{2}$ in terms of $r$, constants of the
motion, and in terms of ${\mbox{\boldmath {\bf L}$\cdot \sigma$} }$ and
${\bf L\cdot S}$ as follows, \begin{eqnarray}
&&v^{2}={\frac{2}{\mu }}E+{\frac{2Gm}{r}}-{\frac{2G({\mbox
{\boldmath{\bf L}$\cdot \sigma$} })}{ c^{2}\mu r^{3}}}\ ,  \label{v2} \\
\dot{r}^{2}=2\frac{E}{\mu }+2\frac{Gm}{r} &&-\frac{L^{2}}{\mu ^{2}r^{2}}+2{
\frac{E({\mbox {\boldmath {\bf L}$\cdot \sigma$} })}{c^{2}m\mu
^{2}r^{2}}}-\frac{2G}{c^{2}\mu r^{3}
}(2{\bf L\cdot S}+{\mbox {\boldmath {\bf L}$\cdot \sigma$} })\ .
\label{radial} \end{eqnarray}
Even though the spin projections of ${\bf L}$ are not conserved, only
the leading-order (conserved) contributions from the
quantities ${\bf L\cdot S_{i}}$ appear in Eqs. (\ref{v2}) and
(\ref{radial}).
 
To complete the description of the motion, we next compute the
evolution of the angle $\gamma$ subtended by the spins ${\bf S_1}$ and ${\bf
S_2}$, further the angles $\kappa_i$ subtended by ${\bf S_i}$ and ${\bf L}$.
In terms of these angles we may express
\begin{equation}  \label{LdotS}
{\bf L\cdot S} = L(S_1 \cos\kappa_1 + S_2\cos\kappa_2)\ ,\qquad {\mbox
{\boldmath {\bf L}$\cdot
\sigma$}} = L(\eta S_1 \cos\kappa_1 + \eta^{-1} S_2\cos\kappa_2) \ .
\end{equation}
 
The variation of the angles $\kappa _{i}$ and $\gamma $ is obtained
from Eqs. (\ref{Sprec}) of the spin precession:
\begin{eqnarray}
(\cos {\bf \skew{44}{\dot}{\kappa}}_{1}) &=&\frac{3G(2+\eta ^{-1})}{
2c^{2}r^{3}}\frac{S_{2}}{L}{\bf L\cdot (\hat{S}_{1}\times \hat{S}_{2})+}
\frac{3G}{c^{2}r^{5}}\frac{S_{2}}{L}{\bf (r\cdot \hat{S}_{2})L\cdot (r\times
\hat{S}_{1})\ ,}  \nonumber \\
(\cos {\bf \skew{44}{\dot}{\kappa}} _{2}{\bf )} &=&{\bf -}\frac{3G(2+\eta )}{
2c^{2}r^{3}}\frac{S_{1}}{L}{\bf L\cdot (\hat{S}_{1}\times \hat{S}_{2})+}
\frac{3G}{c^{2}r^{5}}\frac{S_{1}}{L}{\bf (r\cdot \hat{S}_{1})L\cdot (r\times
\hat{S}_{2})\ ,}  \label{angprec} \\
(\cos {\bf \skew{25}{\dot}{\gamma}} {\bf )} &=&\frac{3G(\eta -\eta ^{-1})}{
2c^{2}r^{3}}{\bf L\cdot (\hat{S}_{1}\times \hat{S}_{2})+}\frac{3G({\bf
r\cdot S}_{2}{\bf -r\cdot S}_{1})}{c^{2}r^{5}}{\bf r\cdot (\hat{S}_{1}\times
\hat{S}_{2})\ .}  \nonumber
\end{eqnarray}
All terms are of order $\epsilon ^{3/2}$,
excepting the first term in $(\cos {\bf \skew{25}{\dot}{\gamma}})$, which is of order $\epsilon$. Thus in Eqs. (\ref{angprec})
we need the mixed products only to leading
order. First we evaluate
\begin{eqnarray}
{\bf L\cdot (\hat{S}_{1}\times \hat{S}_{2})} &=&\mu \left[ ({\bf r\cdot \hat{
S}_{1}})({\bf v\cdot \hat{S}_{2}})-({\bf r\cdot \hat{S}_{2}})({\bf v\cdot
\hat{S}_{1}})\right] \ , \nonumber \\
{\bf L\cdot (r\times \hat{S}_{i})} &=&\mu r\left[ r({\bf v\cdot \hat{S}_{i}}
)-\dot{r}({\bf r\cdot \hat{S}_{i}})\right] \ .  \label{mixed}
\end{eqnarray}
The scalar products may be expressed by use of the angles $\psi $ and $\psi
_{i}$, subtended by the node line ({\sl cf.} Fig.1 in {\bf II}) with the
momentary position vector and the respective projections of the spins in the
plane of orbit. These expressions are:
\begin{eqnarray}
{\bf r\cdot \hat{S}_{i}} &=&r\sin \kappa _{i}\cos (\psi -\psi _{i})\ ,
\nonumber \\
{\bf v\cdot \hat{S}_{i}} &=&\dot{r}\sin \kappa _{i}\cos (\psi -\psi _{i})-
\frac{L}{\mu r}\sin \kappa _{i}\sin (\psi -\psi _{i})\ .  \label{rsvs}
\end{eqnarray}
We have used here the time derivative of the
angle $\psi $ {\em i.e.,} $\dot{\psi}=L/\mu r^{2}$.
 
The angles $\psi_i$ do not independently vary. This can be
seen by noting that the node line is orthogonal both to the total angular
momentum ${\bf J}$ and to the Newtonian orbital momentum ${\bf L_N}$. The
projection of the total spin on the node line yields
\begin{equation}  \label{ang1}
S_1\sin\kappa_1\cos\psi_1+S_2\sin\kappa_2\cos\psi_2=0\ .
\end{equation}
Hence we may express both angles $\psi_i$ in terms of, say, $
\Delta\psi=\psi_2-\psi_1$. Furthermore, from the spherical cosine
identity the
angle $\Delta\psi$ is determined by the angles $\kappa_i$ and $\gamma$
(Fig.1):
\begin{equation}  \label{ang2}
\cos\gamma=\cos\kappa_1\cos\kappa_2+\cos\Delta\psi\sin\kappa_1\sin\kappa_2 \
.
\end{equation}
 
\begin{figure}[htb]
\epsfysize=7cm
\centerline{\hfill
\epsfbox{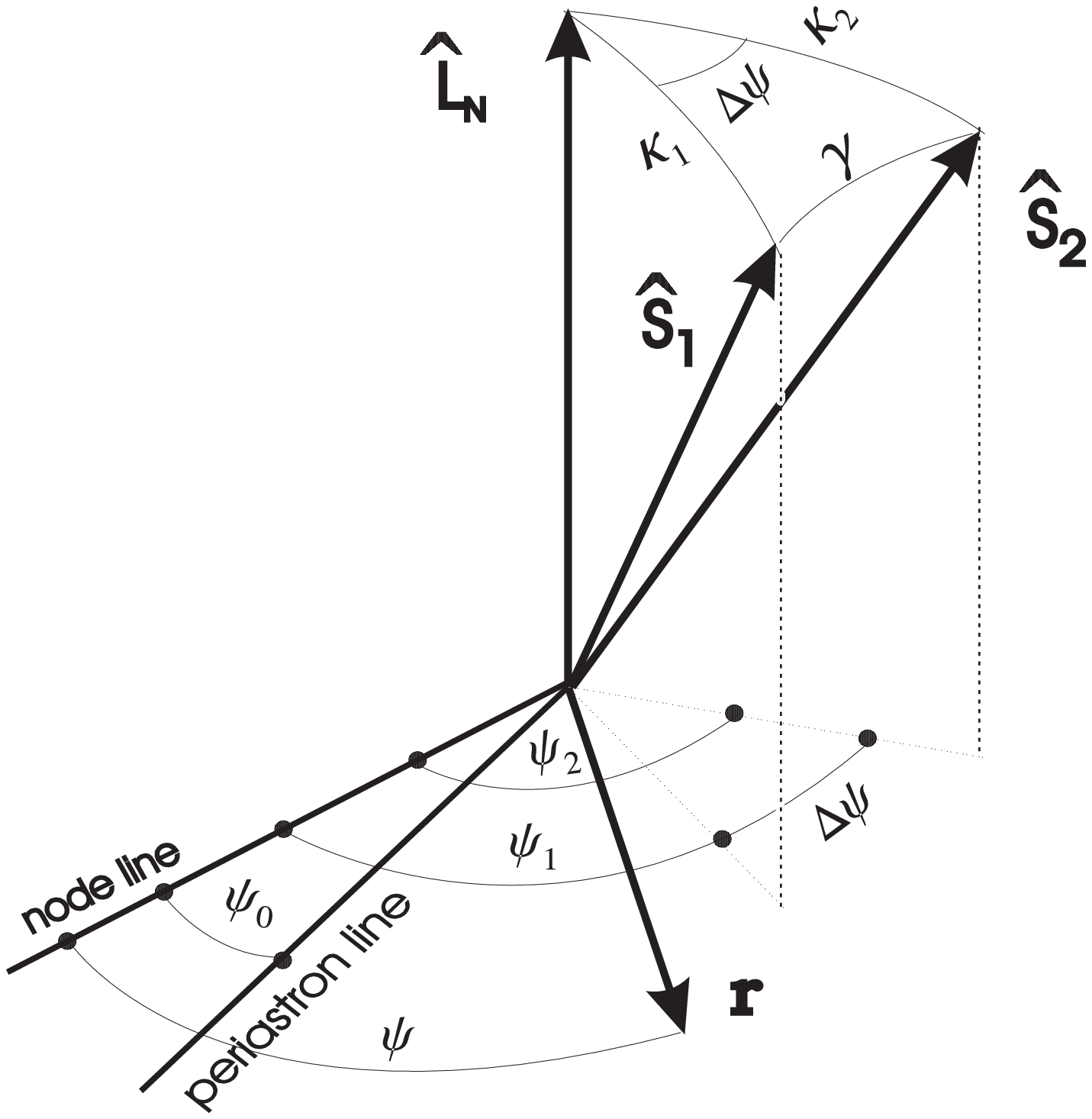}\hfill}
\caption{The angles subtended by ${\bf \hat L_N}$ and ${\bf \hat S_i}$
, and the difference $\Delta\psi$ of the angles subtended by
the node line and the spin projections on the plane of orbit.
To leading order, the Newtonian angular momentum ${\bf L_N=L}$. These angles
are related by the spherical triangle identity. $\psi$ and $\psi_0$ are the
respective angles of the position ${\bf r}$ and the direction of the
periastron with the node line.}
\end{figure}
 
Thus, in terms of the angles $\kappa _{i}$ and $\Delta \psi $ the
expressions (\ref{mixed}), to leading order, take the form:
\begin{eqnarray}
{\bf L\cdot (\hat{S}_{1}\times \hat{S}_{2})} &=&L\sin \kappa _{1}\sin \kappa
_{2}\sin \Delta \psi ,  \nonumber \\
{\bf L\cdot ({r}\times \hat{S}_{i})} &=&\ -Lr\sin \kappa _{i}\sin \left(
\psi -\psi _{i}\right) .  \label{mixed1}
\end{eqnarray}
 
The last mixed product in (\ref{angprec}), again to leading order, works out
as
\begin{equation}
{\bf r\cdot (\hat{S}_{1}\times \hat{S}_{2})}=r[\cos \kappa _{1}\sin \kappa
_{2}\sin (\psi -\psi _{2})-\cos \kappa _{2}\sin \kappa _{1}\sin (\psi -\psi
_{1})]\ .  \label{mixed2}
\end{equation}
 It should be stressed that our treatment of the angles
is fully coordinate invariant.

We proceed next with the parametrization of the orbit. Following the
approach described in Sec. III of {\bf II}, the {\sl eccentric anomaly
parameter} $\xi $ is introduced by
\begin{equation}
r=-{\frac{Gm\mu }{2E}}+{\frac{G\mu (2{\bf L\cdot S+L\cdot}
{\mbox {\boldmath $\sigma$ } })}{
c^{2}L^{2}}}+\left[ {\frac{A_{0}}{2E}}+{\frac{G^{2}m\mu ^{2}(2
{\bf L\cdot S+L\cdot}
{\mbox{\boldmath$\sigma$} })}{c^{2}L^{2}A_{0}}}-{\frac{E({\mbox {\boldmath
{\bf L}$\cdot \sigma$ } })}{ c^{2}m\mu A_{0}}}\right] \cos \xi \ ,  \label{xi}
\end{equation}
where $A_{0}$ is the length of the Runge-Lenz vector to the zeroth order in
the spin:
\begin{equation}
A_{0}=\left( G^{2}m^{2}\mu ^{2}+{\frac{2EL^{2}}{\mu }}\right) ^{1/2}\ .
\label{a0}
\end{equation}
 
As shown in {\bf II}, the orbital period is given by the integral of $
dt/d\xi $ from 0 to $2\pi$,
\begin{equation}
T=2\pi{\frac{Gm\mu^3}{(-2\mu E)^{\frac{3}{2}}}} \ .  \label{period}
\end{equation}
 
The {\sl true anomaly parameter} $\chi $ with properties explained in {\bf I}
(and {\bf II}) is introduced similarly,
\begin{eqnarray}
r=\frac{L^{2}}{\mu (Gm\mu +A_{0}\cos \chi )} &+&\frac{2G(2
{\bf L\cdot
S+L\cdot }{\mbox{\boldmath $\sigma$} })}{c^{2}L^{2}A_{0}}\ \frac{A_{0}(2G^{2}m^{2}\mu
^{3}+EL^{2})+Gm\mu (2G^{2}m^{2}\mu ^{3}+3EL^{2})\cos \chi }{(Gm\mu
+A_{0}\cos \chi )^{2}}  \nonumber \\
&-&\frac{2E({\mbox {\boldmath {\bf L}$\cdot \sigma$} })}{c^{2}m\mu ^{2}A_{0}}\
\frac{Gm\mu
^{2}A_{0}+(G^{2}m^{2}\mu ^{3}+EL^{2})\cos \chi }{(Gm\mu +A_{0}\cos \chi )^{2}
}\ .  \label{chi}
\end{eqnarray}
This yields
\begin{mathletters}
\begin{equation}
{\frac{dt}{d\chi }}={\frac{1}{\dot{r}}}{\frac{dr}{d\chi }}={\frac{\mu r^{2}}{
L}}\left\{ 1-{\frac{1}{c^{2}mL^{4}}}\left[ (2{\bf L\cdot
S+L\cdot }{\mbox {\boldmath $\sigma$} }
)Gm\mu ^{2}(3Gm\mu +A_{0}\cos \chi )-EL^{2}({\mbox {\boldmath {\bf L}$\cdot
\sigma$} })\right] \right\} .  \label{tchi}
\end{equation}
 
$\ $The leading order expressions of the angle variable $\psi $ and of $\dot{
r}$ are also needed in parameterized form:
\end{mathletters}
\begin{equation}
\psi =\psi _{0}+\chi \ ,\qquad \dot{r}={\frac{A_{0}}{L}}\sin \chi \ .
\label{psirdot}
\end{equation}
With (\ref{chi}) and (\ref{psirdot}), all expressions we are interested in
are expressed in terms of the true anomaly parameter $\chi $. We integrate
these expressions as follows,
\begin{equation}
\int_{0}^{T}F(t)dt=\int_{0}^{2\pi }F(\chi ){\frac{dt}{d\chi }}\ d\chi \ .
\end{equation}
The integration is carried out by computing the residues enclosed in the
circle $\zeta =e^{i\chi }$. As a rule, we find that there is only one
pole\footnote{We were initially unaware of the conditions under which
the true anomaly parameter has this property. A systematic approach
has recently been developed \cite{param} to the
parametrizations of the perturbed Kepler motion. By its use,
we can now prove that the poles of these integrands
are in the origin of the complex parameter plane.},
at $\zeta =0$.
 
For a first application, we compute the average rate of change of the
angles.
\begin{eqnarray}
\label{angprecave}
\left\langle (\cos {\bf \skew{44}{\dot}\kappa} _{1}{\bf )}\ \right\rangle
{\bf \,} &=&\frac{3G(1+\eta ^{-1})(-2\mu E)^{3/2}S_{2}}{2c^{2}L^{3}}
\sin \kappa _{1}\sin \kappa _{2}\sin \Delta \psi {\bf \ ,}  \nonumber \\
\left\langle (\cos {\bf \skew{44}{\dot}\kappa} _{2}{\bf )}\ \right\rangle
{\bf \,} &=&{\bf -}\frac{3G(1+\eta )(-2\mu E)^{3/2}S_{1}}{2c^{2}L^{3}}\sin \kappa
_{1}\sin \kappa _{2}\sin \Delta \psi {\bf \ ,}  \\
\left\langle (\cos {\bf \skew{27}{\dot}\gamma} {\bf )\,}\ \right\rangle &=&
\frac{3G(-2\mu E)^{3/2}}{2c^{2}L^{2}}\left( \eta -\eta ^{-1}+\frac{
S_{1}}{L}\cos \kappa _{1}-\frac{S_{2}}{L}\cos \kappa _{2}\right) \sin \kappa
_{1}\sin \kappa _{2}\sin \Delta \psi {\bf \ .}  \nonumber
\end{eqnarray}
    The secular changes do not vanish. This should be contrasted
with the behavior of the one-spin system, presented in {\bf II}, where
the angular momenta are frozen in a rigidly rotating parallelogram, and
the relative angles are constant.
 
  We may express the secular changes of the angles in terms of the
semimajor axis $a=(Gm\mu)/(-2E)$ and eccentricity $e$ given by
$1-e^2=(-2EL^2)/(G^2m^2\mu^3)$. The Keplerian values are
sufficient to the accuracy needed here. We then get
\begin{eqnarray}
\label{angprecave2}
\left\langle (\cos {\bf \skew{44}{\dot}\kappa} _{1}{\bf )}\ \right\rangle
{\bf \,} &=&\frac{3G(1+\eta ^{-1})S_{2}}{2c^{2}a^{3}(1-e^2)^{3/2}}
\sin \kappa _{1}\sin \kappa _{2}\sin \Delta \psi {\bf \ ,}  \nonumber \\
\left\langle (\cos {\bf \skew{44}{\dot}\kappa} _{2}{\bf )}\ \right\rangle
{\bf \,} &=&{\bf -}\frac{3G(1+\eta )S_{1}}{2c^{2}a^{3}(1-e^2)^{3/2}}\sin \kappa
_{1}\sin \kappa _{2}\sin \Delta \psi {\bf \ ,}  \\
\left\langle (\cos {\bf \skew{27}{\dot}\gamma} {\bf )\,}\ \right\rangle &=&
\frac{3G\left[(\eta -\eta ^{-1}) \mu\left(Gma(1-e^2)\right)^{1/2}+S_{1}\cos \kappa _{1}-S_{2}\cos \kappa _{2}\right]}{2c^{2}a^{3}(1-e^2)^{3/2}} \sin \kappa
_{1}\sin \kappa _{2}\sin \Delta \psi {\bf \ .}  \nonumber
\end{eqnarray}
 
  In the particular case when the masses and spin magnitudes are equal,
$m_1=m_2$ and $S_1=S_2$ and with the appropriate change of notation,
these equations
agree with Apostolatos'\cite{Apost} Eqs. (4) up to a constant multiplier
$1/(1-e^2)^{3/2}$ of the time $t$. Thus the eccentricity $e$ of the
orbit speeds up the evolution of the angular variables $\kappa_i$ and
$\gamma$. In the generic case, with arbitrary values of the masses
and spin magnitudes, the scaling factor is still present in the
denominators of Eqs. (\ref{angprecave}), but no corresponding term in
the numerators. We thus find that the eccentricity of the orbit
accelerates the evolution of the spin directions.
 
\section{Instantaneous radiative losses}
 
We obtain the instantaneous losses of the constants of motion $E$ and $L$
and the angles $\kappa _{i}$ and $\gamma $ using Kidder's \cite{Kidder}
results and the Burke-Thorne potential \cite{BT}. The variation of the
remaining angles follows by the relations (\ref{ang1}) and (\ref{ang2}).
 
First we find the radiative spin losses. To lowest order the
radiation-reaction potential is the Burke-Thorne potential \cite{BT,Blanchet}
:
\begin{equation}
V=-{\frac{G}{5c^{5}}}I_{\mu\nu}^{(5)}y_{\mu}y_{\nu}\ ,  \label{BTpot}
\end{equation}
where $I_{\mu\nu}^{(5)}$ is the fifth time derivative of the system's
quadrupole-moment tensor and $y_{\mu}$ are Cartesian coordinates centered on
the spinning body.
 
The radiative spin loss evaluated in {\bf II }holds for each of the spins in
its system of principal axes of inertia:
\begin{equation}
\frac{1}{S_{i}}\frac{d\left( {\bf S_{i}}\right) _{\mu}}{dt}={\frac{2G}{
5c^{5}\Omega _{i}}}\left( {\frac{\Theta _{i}}{\Theta _{i}^{\prime }}}
-1\right) \epsilon _{\mu\nu\rho}I_{\nu\sigma}^{(5)}( {\bf \hat{S}_{i}})
_{\rho}( {\bf \hat{S}_{i}}) _{\sigma}.  \label{Sdirdot}
\end{equation}
Here $\Theta _{i}$ and $\Theta _{i}^{\prime }$ are the principal moments of
inertia and $\Omega _{i}$ is the angular velocity of the $i^{th}$ spinning
axisymmetric body. These quantities are related by $S_{i}=\Theta
_{i}^{\prime }\Omega _{i}.$
 
Two important properties of the radiative spin loss $d{\bf S_{i}}/dt$ emerge
from (\ref{Sdirdot}):
 
$(i)$ they are of $2^{nd}$ post-Newtonian order and
 
$(ii)$ are each perpendicular to the respective spins ${\bf S_{i}}$.
 
Next we consider the energy loss $dE/dt$ and total
angular momentum loss $d{\bf J}/dt$.
These have been computed by Kidder using the
Blanchet-Damour-Iyer formalism\cite{BDI}. Keeping the Newtonian
and spin-orbit terms in the expressions, we have
\begin{eqnarray}
{\frac{dE}{dt}}=- &&{\frac{8G^{3}m^{2}\mu ^{2}}{15c^{5}r^{4}}}\,(12v^{2}-11
\dot{r}^{2})  \label{dEdt} \\
- &&{\frac{8G^{3}m\mu }{15c^{7}r^{6}}}\,\left[ ({\bf L_{N}}\!\cdot \!{\bf S}
)\left( 27\dot{r}^{2}-37v^{2}-12{\frac{Gm}{r}}\right) +({\bf L_{N}}\!\cdot \!
{\mbox{\boldmath $\sigma$} })\left( 51\dot{r}^{2}-43v^{2}+4{\frac{Gm}{r}}\right) \right]\ ,
\nonumber \\
{\frac{d{\bf J}}{dt}}=- &&{\frac{8}{5}}{\frac{G^{2}m\mu }{c^{5}r^{3}}}{\bf L
_{N}}\left( -3\dot{r}^{2}+2v^{2}+2{\frac{Gm}{r}}\right) -{\frac{4}{5}}{\frac{
G^{2}\mu ^{2}}{c^{7}r^{3}}}\Biggl\{-{\frac{2}{3}}{\frac{Gm}{r}}(\dot{r}
^{2}-v^{2})({\bf S}-\mbox {\boldmath$\sigma$}
)-\dot{r}{\frac{Gm}{3r^{2}}}{\bf
r}\times [{\bf v} \times (7{\bf S}+5\mbox{\boldmath$\sigma$} )]
\nonumber \\
+ &&{\frac{Gm}{r^{3}}}{\bf r}\times \left[ ({\bf r}\times {\bf S})\left( 6
\dot{r}^{2}-{\frac{17}{3}}v^{2}+2{\frac{Gm}{r}}\right) +({\bf r}\times
{\mbox{\boldmath$
\sigma$} })\left( 9\dot{r}^{2}-8v^{2}-{\frac{2}{3}}{\frac{Gm}{r}}\right)
\right]  \nonumber \\
+ &&{\frac{\dot{r}}{r}}{\bf v}\times \left[ ({\bf r}\times {\bf S})\left( -30
\dot{r}^{2}+24v^{2}+{\frac{29}{3}}{\frac{Gm}{r}}\right) +5({\bf r}\times
{\mbox{\boldmath $\sigma$} })\left( -5\dot{r}^{2}+4v^{2}+{\frac{5}{3}}{\frac{Gm}{r}}\right)
\right]  \label{dJdt} \\
+ &&{\bf v}\times \left[ ({\bf v}\times {\bf S})\left( 18\dot{r}^{2}-12v^{2}-
{\frac{23}{3}}{\frac{Gm}{r}}\right) +({\bf v}\times {\mbox{\boldmath $\sigma$} })\left( 18
\dot{r}^{2}-{\frac{35}{3}}v^{2}-9{\frac{Gm}{r}}\right) \right]  \nonumber \\
+ &&{\frac{{\bf L_{N}}}{\mu ^{2}r^{2}}}\left[ ({\bf L_{N}}\cdot {\bf S}
)\left( 30\dot{r}^{2}-18v^{2}-{\frac{92}{3}}{\frac{Gm}{r}}\right) +({\bf
L_{N}}\cdot {\mbox{\boldmath $\sigma$} })\left( 35\dot{r}^{2}-19v^{2}-{\frac{71}{3}}{\frac{
Gm}{r}}\right) \right] \Biggr\}\ .  \nonumber
\end{eqnarray}
 
The loss $dL/dt={\bf \hat{L}\cdot }d{\bf J}/dt$ in the magnitude of the
orbital angular momentum follows from (\ref{dJdt}) as the spin-orbit terms
in $d{\bf J}/dt$ do not receive contributions from the radiative spin
losses, {\it cf.} property $(i)$. All mixed vector products in
(\ref{dJdt}), when projected to ${\bf \hat L}$,
can be converted to one of $({\bf L\cdot S})$ and $({\mbox{\boldmath$
{\bf L}\cdot \sigma$} })$.
The first, Newtonian, term is expressed by use of Eqs. (\ref{Lvect}) and
(\ref{LSO}). Using then Eqs. (\ref{v2}), (\ref{radial}) and (\ref{LdotS}) we
obtain purely radial expressions for the instantaneous losses:
\begin{eqnarray}
{\frac{dE}{dt}} &=&-\frac{8\ G^{3}m^{2}}{15c^{5}r^{6}}\left( 2\mu
Er^{2}+2Gm\mu ^{2}r+11L^{2}\right) +\frac{8G^{3}mL}{15c^{7}\mu r^{8}}
\label{instlossE} \\
&&\times \Bigl[\left( 20\mu Er^{2}-12Gm\mu ^{2}r+27L^{2}\right) (S_{1}\cos
\kappa _{1}+S_{2}\cos \kappa _{2})  \nonumber \\
&&+\left( 6\mu Er^{2}-18Gm\mu ^{2}r+51L^{2}\right) (\eta S_{1}\cos \kappa
_{1}+\eta ^{-1}S_{2}\cos \kappa _{2})\Bigr]  \ ,\nonumber \\
{\frac{dL}{dt}} &=&\frac{8G^{2}mL}{5c^{5}\mu r^{5}}\left( 2\mu
Er^{2}-3L^{2}\right) +\frac{8G^{2}}{15c^{7}\mu ^{2}r^{7}}  \label{instlossL}
\\
&&\times \Bigl\{\mu r\left[ 12Gm\mu ^{2}Er^{2}+3\left( G^{2}m^{2}\mu
^{3}+6EL^{2}\right) r-11Gm\mu L^{2}\right] (S_{1}\cos \kappa _{1}+S_{2}\cos
\kappa _{2})  \nonumber \\
&&+\left[ 2\mu ^{2}E^{2}r^{4}+12Gm\mu ^{3}Er^{3}+3\mu (G^{2}m^{2}\mu
^{3}+5EL^{2})r^{2}-5Gm\mu ^{2}L^{2}r+15L^{4}\right] (\eta S_{1}\cos \kappa
_{1}+\eta ^{-1}S_{2}\cos \kappa _{2})\Bigr\}\ .  \nonumber
\end{eqnarray}
 
We also need the projections of the orbital angular momentum loss in the
directions of the spins for evaluating the radiative change in the angle
variables. The foregoing considerations yield these projections, after a
straightforward but cumbersome computation, in the form
\begin{equation}
{\frac{d{\bf L}}{dt}}{\bf \cdot \hat{S}_{i}}=\frac{dL}{dt}\cos \kappa _{i}+
\frac{2G^{2}}{15c^{7}\mu ^{2}r^{7}}\sum_{j=1}^2\left(
2p_{ij} + \eta^{3-2j}q_{ij}\right)
S_{j}\sin \kappa _{i}\sin \kappa _{j}              \label{instlossLproj}
\end{equation}
with
\begin{eqnarray}
p_{ij}= &&9\mu L(-2\mu Er^{2}+5L^{2})r\dot{r}\sin (\psi _{j}-\psi _{i})
\label{intlossLproj3} \nonumber \\
+ &&[12Gm\mu ^{3}Er^{3}+3\mu \left( G^{2}m^{2}\mu ^{3}+18EL^{2}\right)
r^{2}+14Gm\mu ^{2}L^{2}r-45L^{4}]\cos (\psi _{j}-\psi _{i}) \nonumber \\
- &&3[4Gm\mu ^{3}Er^{3}+\mu \left( G^{2}m^{2}\mu ^{3}+6EL^{2}\right)
r^{2}+Gm\mu ^{2}L^{2}r+3L^{4}]\cos (2\psi -\psi _{i}-\psi _{j}) \nonumber \\
+ &&3\mu L(6\mu Er^{2}+2Gm\mu ^{2}r-3L^{2})r\dot{r}\sin (2\psi -\psi
_{i}-\psi _{j}) \ ,  \\
q_{ij}= &&15\mu L(-2\mu Er^{2}+5L^{2})r\dot{r}\sin (\psi _{j}-\psi _{i})
\label{intlossLproj4}  \nonumber \\
+ &&[4\mu ^{2}E^{2}r^{4}+24Gm\mu ^{3}Er^{3}+6\mu \left( G^{2}m^{2}\mu
^{3}+15EL^{2}\right) r^{2}+29Gm\mu ^{2}L^{2}r-75L^{4}]\cos (\psi _{j}-\psi
_{i})  \nonumber \\
- &&[4\mu ^{2}E^{2}r^{4}+24Gm\mu ^{3}Er^{3}+2\mu \left( 3G^{2}m^{2}\mu
^{3}+13EL^{2}\right) r^{2}+3Gm\mu ^{2}L^{2}r+15L^{4}]\cos (2\psi -\psi
_{i}-\psi _{j})  \nonumber \\
+ &&\mu L(34\mu Er^{2}+12Gm\mu ^{2}r-15L^{2})r\dot{r}\sin (2\psi -\psi
_{i}-\psi _{j}).
\end{eqnarray}
These expressions depend on $r$ as well as on the angular variable
$\psi$ and on the derivative $\dot r$.
 
Thus the spin projection of the instantaneous change in ${\bf \hat{L}}$ is
\begin{equation}
{\frac{d{\bf \hat{L}}}{dt}}{\bf \cdot \hat{S}_{i}} = \frac{1}{L}\left( {
\frac{d{\bf L}}{dt}}{\bf \cdot \hat{S}_{i}}-\frac{dL}{dt}\cos \kappa
_{i}\right)
= \frac{2G^{2}}{15c^{7}\mu ^{2}Lr^{7}}\sum_{j=1}^2\left(
2p_{ij} + \eta^{3-2j}q_{ij}\right)
S_{j}\sin \kappa _{i}\sin \kappa _{j}
\label{instlossLdirproj} \ .
\end{equation}
 
Due to property $(ii)$ the instantaneous change in ${\bf \hat{S}_{i}}$ equals
the right hand side of (\ref{Sdirdot}) and its projection to ${\bf \hat{L}}$
yields:
\begin{eqnarray}
{\bf \hat{L}}\cdot \frac{d{\bf \hat{S}_{i}}}{dt} &=&\frac{2G^{2}m}{5c^{5}\mu
^{2}r^{7}\Omega _{i}}\left( \frac{\Theta _{i}}{\Theta _{i}^{\prime }}
-1\right) \sin ^{2}\kappa _{i}  \label{instlossSdirproj} \\
&&\times \left[ 4L(-18E\mu r^{2}-20Gm\mu ^{2}r+15L^{2})\cos (2\psi -2\psi
_{i})-\mu r\dot{r}(12E\mu r^{2}+20Gm\mu ^{2}r+45L^{2})\sin (2\psi -2\psi
_{i})\right] .  \nonumber
\end{eqnarray}
 
We are now in position to compute the radiative losses of $\cos
\kappa _{i}$ :
\begin{equation}
\frac{d\cos \kappa _{i}}{dt}={\frac{d}{dt}}({\bf \hat{L}}\cdot {\bf \hat{S}
_{i}})={\frac{d{\bf \hat{L}}}{dt}}{\bf \cdot \hat{S}_{i}}+{\bf \hat{L}}\cdot
\frac{d{\bf \hat{S}_{i}}}{dt}\ .  \label{instlosska}
\end{equation}
 
Finally we find from Eq. (\ref{Sdirdot}) the radiative change in the
angle $ \gamma $ subtended by the spin vectors:
\begin{eqnarray}
\frac{d\cos\gamma }{dt}=- &&\frac{2G^{2}m}{5c^{5}\mu^{2}r^{7}}\sum_{i\neq j}
\frac{1}{\Omega _{i}}\left( \frac{\Theta _{i}}{\Theta _{i}^{\prime }}
-1\right) \sin \kappa _{i}  \nonumber \\
\times \Bigl\{\mu r &&\dot{r}\Bigl[(12E\mu r^{2}+20Gm\mu
^{2}r+45L^{2})\left( \sin \kappa _{i}\cos \kappa _{j}\sin (2\psi -2\psi
_{i})-\cos \kappa _{i}\sin \kappa _{j}\sin (2\psi -\psi _{i}-\psi
_{j})\right)  \nonumber \\
&&+(12E\mu r^{2}+20Gm\mu ^{2}r-15L^{2})\cos \kappa _{i}\sin \kappa _{j}\sin
(\psi _{j}-\psi _{i})\Bigr]  \label{instlossga} \\
+4 &&L(18E\mu r^{2}+20Gm\mu ^{2}r-15L^{2})\left( \sin \kappa _{i}\cos \kappa
_{j}\cos (2\psi -2\psi _{i})-\cos \kappa _{i}\sin \kappa _{j}\cos (2\psi
-\psi _{i}-\psi _{j})\right) \Bigr\}.  \nonumber
\end{eqnarray}
 
We emphasize that in spite of the fact that the radiative spin losses $d{\bf
S_{i}}/dt$ are of $2^{nd}$ post-Newtonian order, they contribute at $
\epsilon ^{3/2}$ order to the instantaneous angular losses.
 
\section{Averaged radiative losses}
 
The instantaneous losses of the constants of motion and the angles subtended
by the orbital and spin angular momenta, Eqs. (\ref{instlossE}) - (\ref
{instlossga}) are in a form suitable for parametrization with the true
anomaly parameter $\chi $, using (\ref{chi}) and (\ref{psirdot}). Then the
averaged losses are computed by the residue theorem, passing to the complex
variable $\zeta =e^{i\chi }$.
 
The averaging procedure yields for the constants of motion:
\begin{eqnarray}
\left\langle {\frac{dE}{dt}}\right\rangle &=&-\frac{G^{2}m(-2E\mu )^{3/2}}{
15c^{5}L^{7}}(148E^{2}L^{4}+732G^{2}m^{2}\mu ^{3}EL+425G^{4}m^{4}\mu ^{6})+
\frac{G^{2}(-2E\mu )^{3/2}}{10c^{7}L^{10}}  \nonumber \\
&\times &\bigl\{(520E^{3}L^{6}+10740G^{2}m^{2}\mu
^{3}E^{2}L^{4}+24990G^{4}m^{4}\mu ^{6}EL^{2}+12579G^{6}m^{6}\mu
^{9})(S_{1}\cos \kappa _{1}+S_{2}\cos \kappa _{2})  \nonumber \\
&+&(256E^{3}L^{6}+6660G^{2}m^{2}\mu ^{3}E^{2}L^{4}+16660G^{4}m^{4}\mu
^{6}EL^{2}+8673G^{6}m^{6}\mu ^{9})(\eta S_{1}\cos \kappa _{1}+\eta
^{-1}S_{2}\cos \kappa _{2})\bigr\}  \label{avelossE} \\
\left\langle {\frac{dL}{dt}}\right\rangle &=&-\frac{4G^{2}m(-2E\mu )^{3/2}}{
5c^{5}L^{4}}(14EL^{2}+15G^{2}m^{2}\mu ^{3})+\frac{G^{2}(-2E\mu )^{3/2}}{
15c^{7}L^{7}}  \nonumber \\
&\times &\bigl\{ (1188E^{2}L^{4}+6756G^{2}m^{2}\mu
^{3}EL^{2}+5345G^{4}m^{4}\mu ^{6})(S_{1}\cos \kappa _{1}+S_{2}\cos \kappa
_{2})  \nonumber \\
&+&(772E^{2}L^{4}+4476G^{2}m^{2}\mu ^{3}EL^{2}+3665G^{4}m^{4}\mu ^{6})(\eta
S_{1}\cos \kappa _{1}+\eta ^{-1}S_{2}\cos \kappa _{2})\bigr\}
\label{avelossL}
\end{eqnarray}
 
For the angle variables, we first find that (\ref{instlossSdirproj}) and (
\ref{instlossga}) average out:
 
\begin{equation}
\left\langle {\bf \hat{L}}\cdot \frac{d{\bf \hat{S}_{i}}}{dt}\right\rangle
=0\ \;,\;\;\qquad \left\langle \frac{d\gamma }{dt}\right\rangle =0.
\label{avelossga}
\end{equation}
Thus, when averaging Eq. (\ref{instlosska}),
only the first term survives, which leaves us with
\begin{eqnarray}   \label{avelosska}
\left\langle \frac{d\kappa _{1}}{dt}\right\rangle &=&\frac{G^{2}(-2E\mu
)^{3/2}}{30c^{7}L^{8}}\bigl\{ (1140E^{2}L^{4}\!\!+\!4164Gm^{2}\mu
^{3}EL^{2}+2285G^{4}m^{4}\mu ^{6})(S_{1}\sin \kappa _{1}+S_{2}\sin \kappa
_{2}\cos \Delta \psi )                    \\
+(884E^{2}L^{4} &+&3264G^{2}m^{2}\mu ^{3}EL^{2}+1795G^{4}m^{4}\mu ^{6})(\eta
S_{1}\sin \kappa _{1}+\eta ^{-1}S_{2}\sin \kappa _{2}\cos \Delta \psi )
\nonumber \\
+6(52E^{2}L^{4} &+&92G^{2}m^{2}\mu ^{3}EL^{2}\!+\!33G^{4}m^{4}\mu
^{6})[S_{1}\sin \kappa _{1}\cos (2\psi _{1}\!-\!2\psi _{0})\!+\!S_{2}\sin
\kappa _{2}\cos (\psi _{1}\!+\!\psi _{2}\!-\!2\psi _{0})]  \nonumber \\
+\!2(238E^{2}L^{4}\!\! &+&\!431G^{2}m^{2}\mu
^{3}EL^{2}\!\!\!+\!\!156G^{4}m^{4}\mu ^{6})[\eta S_{1}\!\sin \kappa _{1}\!\cos
(2\psi _{1}\!\!-\!2\psi _{0})\!\!+\!\eta ^{-1}\!S_{2}\sin \kappa _{2}\cos
(\psi _{1}\!\!+\!\psi _{2}\!\!-\!2\psi _{0})]\bigr\}\,.  \nonumber
\end{eqnarray}
The corresponding expression $\left\langle {d\kappa _{2}}/{dt}\right\rangle $
is obtained by swapping the indices $1\leftrightarrow 2$ in
(\ref{avelosska}) and substituting $\eta \leftrightarrow \eta ^{-1}$.
 
  For completeness we give the radiative evolution equations also in
terms of orbital elements suited for our perturbative treatment.
The generalized semimajor axis $a$ and eccentricity $e$
are introduced by $r_{{}_{{}^{max}_{min}}}=a(1\pm e)$. (The turning points
$r_{{}_{{}^{max}_{min}}}$ follow from (\ref{xi}) inserting $\xi=\pi$ and
$0$ respectively.)
\begin{eqnarray} \label{a1me2}
a=&-&{Gm\mu \over 2E}\Bigl[1-{2E(2{\bf L\cdot S}+{\mbox{\boldmath{\bf L}$\cdot
\sigma$}})
                              \over c^2mL^2}\Bigr]
  \ ,\nonumber\\
1-e^2=-{2EL^2\over G^2m^2\mu^3}\Bigl[1&
          + &{4(2{\bf L\cdot S}+{\mbox{\boldmath{\bf L}$\cdot \sigma$}
})\over c^2mL^4}
  (G^2m^2\mu^3+EL^2)-{2E({\mbox{\boldmath{\bf L}$\cdot\sigma$}})\over
c^2mL^2}
                \Bigr]\ .
\end{eqnarray}
The inverse relations are
\begin{eqnarray} \label{el2}
E=&-&{Gm\mu \over 2\,a}\Bigl[1+{G^{1/2}\over c^2m^{1/2}}\,
                  {(2+\eta)S_1\cos\kappa_1
                 + (2+\eta^{-1})S_2\cos\kappa_2\over
                     a^{3/2}(1-e^2)^{1/2}}\Bigr]\ ,\nonumber\\
L^2=Gm\mu^2 a(1&-&e^2)\Bigl\{1-{2G^{1/2}\over
  c^2m^{1/2}a^{3/2}(1-e^2)^{3/2}}\left[
          S_1\cos\kappa_1(3+e^2+2\eta)
         +S_2\cos\kappa_2(3+e^2+2\eta^{-1})\right]
            \Bigr\}\ .
\end{eqnarray}
The first terms on the right-hand sides represent the Keplerian
approximation. We emphasize here that the expressions of the
angular precessions (\ref{angprecave2}) are unchanged if we rewrite
them in terms of $a$ and $e$ defined above.
 
  A straightforward computation yields the averaged radiation losses of
$a$, $e$ and $\kappa_i$:
\begin{eqnarray}   \label{avelossa}
\left\langle \frac{da}{dt}\right\rangle =
&-&\frac{2G^3m^2\mu (37e^4+292e^2+96)}{15c^5a^3(1-e^2)^{7/2}} \nonumber\\
&+&\frac{G^{7/2}m^{3/2}\mu}{15c^7a^{9/2}(1-e^2)^5} \Bigl\{
 (363e^6+3510e^4+7936e^2+2128) (S_1\cos\kappa_1+S_2\cos\kappa_2)  \\
&+&(291e^6+4224e^4+7924e^2+1680)
  (\eta S_1\cos\kappa_1+\eta^{-1}S_2\cos\kappa_2)  \Bigr\} \nonumber\ ,
\end{eqnarray}
\begin{eqnarray}   \label{avelosse}
\left\langle \frac{de}{dt}\right\rangle =
&-&\frac{G^3m^2\mu e\,(121e^2+304)}{15c^5a^4(1-e^2)^{5/2}} \nonumber\\
&+&\frac{G^{7/2}m^{3/2}\mu e}{30c^7a^{11/2}(1-e^2)^4} \Bigl\{
 (1313e^4+5592e^2+7032) (S_1\cos\kappa_1+S_2\cos\kappa_2) \\
&+& (1097e^4+6822e^2+6200)
  (\eta S_1\cos\kappa_1+\eta^{-1}S_2\cos\kappa_2) \Bigr\} \nonumber\ ,
\end{eqnarray}
\begin{eqnarray}        \label{avelosska1}
\left\langle \frac{d\kappa_1}{dt}\right\rangle &=
&\frac {G^{7/2}m^{3/2}\mu}{30c^7a^{11/2}(1-e^2)^4} \Bigl\{
 (285e^4+1512e^2+488)
   (S_1\sin\kappa_1+S_2\sin\kappa_2\cos\Delta\psi) \nonumber\\
&+& (221e^4+1190e^2+384)
 (\eta S_1\sin\kappa_1+\eta^{-1}S_2\sin\kappa_2\cos\Delta\psi) \\
&+& (156e^4+240e^2)
  \bigl[S_1\sin\kappa_1\cos(2\psi_1-2\psi_0)
    +S_2\sin\kappa_2\cos(\psi_1+\psi_2-2\psi_0)\bigr] \nonumber\\
&+& (119e^4+193e^2)
 \bigl[\eta S_1\sin\kappa_1\cos(2\psi_1-2\psi_0)
   +\eta^{-1}S_2\sin\kappa_2\cos(\psi_1+\psi_2-2\psi_0)\bigr]\Bigr\}
\nonumber\ .
\end{eqnarray}
The substitutions following (\ref{avelosska}) should be carried out once
more to get the secular variation of $\kappa_2$.
 
 The averaged losses (\ref{avelossE}), (\ref{avelossL}), (\ref{avelossga})
and (\ref{avelosska}) (and the similar expression for $\left\langle {d\kappa
_{2}}/{dt}\right\rangle $), together with the algebraic relations (\ref{ang1})
and (\ref{ang2}) provide a complete description of the radiative evolution
of the binary system, to $\epsilon ^{3/2}$ order in terms of
radiative losses. On the other hand the radiative changes of the
angles $\kappa_i$ are of $\epsilon ^{5/2}$ relative order
compared to their secular changes given in (\ref{angprecave2}).
An alternative set of
evolution equations, in terms of the orbital elements $a$, $e$ and
$\kappa_i$ is provided by Eqs. (\ref{angprecave2}),
(\ref{avelossa}), (\ref{avelosse}) and (\ref{avelosska1}).
It is remarkable that all
the contributions from the radiative losses of the spins, present in the
instantaneous losses average to zero.
 
\section{Concluding Remarks}
 
Our description of the smoothed evolution of a spinning binary system
opens up the possibility to study the modulation of the
gravitational wave forms induced by the eccentricity.
Investigations of the smoothed evolution of circular orbits have
already been presented in several papers (\cite{ACST}, \cite{Apost}
and \cite{deFelice}). The detailed analysis of the
angular evolution equations is a subtle issue. An immediate effect that
follows from Eqs. (\ref{angprecave}) of Sec. II is the acceleration of
the evolution of spin orientations with increasing eccentricity.
 
The radiative losses of both the dynamical quantities $E$ and $L$ and of
the angular variables $\kappa_i$ and $\gamma$ subtended
by the angular momenta were given here up to $\epsilon^{3/2}$ order
compared to the leading order losses. Among them the angular
losses of the angles $\kappa_i$ are of $\epsilon^{5/2}$ order
beyond the secular spin-orbit terms given in (\ref{angprecave2}).
Our previous results in {\bf I} and {\bf II} are particular cases
of the present radiative loss equations.
The one-spin limit arises by $S_2\to 0$ and $\psi_1=\pi/2$ (the latter
relation stems from (\ref{ang1})). For the Lense-Thirring case the
additional limit $\eta\to 0$ has to be taken.
We would also like to point out the agreement of
the energy and orbital momentum losses with computations
in a different, noncovariant {\em SSC}. There the energy $E$ and
orbital momentum $L$ were derived from a different action, but their
radiative losses, given by Rieth and Sch\"afer \cite{RS} coincide with
those in our Eqs. (\ref{avelossE}) and (\ref{avelossL}).
However, in our approach, the previously unknown radiative evolution
of the angles $\kappa_i$ and $\gamma$, characterizing the geometry of
the binary system, could readily been obtained.
 
\section{Acknowledgments}
 
This work has been supported by OTKA no. T017176 and D23744 grants. The
algebraic package REDUCE was used for checking our computations.

\end{document}